\documentclass[allclo]{FBSart}
\usepackage{amsfonts}
\usepackage{amssymb}
\usepackage{epsfig}

\def\be{\begin{equation}}
\def\ee{\end{equation}}

\newcommand{\nn}{\nonumber}

\newcommand{\ba}{\begin{eqnarray}}
\newcommand{\ea}{\end{eqnarray}}
\newcommand{\Eq}[1]{Eq.~(\ref{#1})}
\newcommand{\Eqs}[1]{Eqs.~(\ref{#1})}

\newcommand{\Ref}[1]{Ref.~\cite{#1}}

\newcommand{\Figs}[1]{Figs.~{\ref{#1}}}

\title{Chiral Extrapolation of Lattice Data for Heavy Meson
Hyperfine Splittings}

\author{X.-H. Guo\instnr{1,2,}\thanks{\textit{E-mail address:} xhguo@bnu.edu.cn},
P.~C. Tandy\instnr{3,}\thanks{\textit{E-mail address:} tandy@cnr2.kent.edu},
A.~W. Thomas\instnr{4,}\thanks{\textit{E-mail address:} awthomas@jlab.org}
}

\instlist{Key Laboratory of Radiation
Beam Technology and Material Modification of National Ministry of
Education, and Institute of Low Energy Nuclear Physics, Beijing
Normal University, Beijing 100875, China
\and Department of Physics and Mathematical Physics, and Special
Research Center for the Subatomic Structure of Matter, Adelaide
University, SA 5005, Australia
\and Center for Nuclear Research, Department
of Physics, Kent State University, Kent, Ohio 44242, USA
\and Jefferson Lab, 12000 Jefferson Avenue, Newport News,
VA 23606, USA}

\runningauthor{X.-H.\,Guo}
\runningtitle{Chiral Extrapolation of Lattice Data}
\begin{document}

\maketitle
\begin{abstract}
We investigate the chiral extrapolation of the lattice data for the
light-heavy meson hyperfine
splittings $D^*-D$ and $B^*-B$ to the physical region for the light quark mass.
The chiral loop corrections providing non-analytic behavior in $m_\pi$ are
consistent with chiral perturbation theory for heavy mesons.  Since chiral
loop corrections tend to decrease the already too low splittings obtained from
linear extrapolation, we investigate two
models to guide the form of the analytic background behavior: the constituent
quark potential model, and the covariant model of QCD based on the ladder-rainbow
truncation of the Dyson-Schwinger equations.   The extrapolated hyperfine
splittings remain clearly below the experimental values even allowing for the
model dependence in the description of the analytic background.
\end{abstract}

\section{\label{sec:intro} Introduction}

In the past few years there has been much progress in lattice gauge theory with
many physical quantities having been calculated. Among them
the hyperfine splittings in the heavy meson systems are of particular
interest. With the aid of nonrelativistic QCD (NRQCD) on the lattice,
the authors of Ref. \cite{hein} reported  three lattice data for the
$qQ$ meson hyperfine splittings $D^*-D$ and $B^*-B$.  These data were
obtained in the unphysical region where $m_q$ corresponds to $m_\pi$ being
larger than
about 680~MeV. With a naive linear extrapolation to the physical
$m_\pi$, the extrapolated hyperfine  splittings are typically 120~MeV for
$D^*-D$, and 32~MeV for $B^*-B$.  These are significantly smaller than the
experimental values 140~MeV and 46~MeV respectively.  The obvious shortcoming
in the naive linear extrapolation is that the nonanalytic terms in the light
quark mass, generated by chiral loops, do not appear.

To correct this, we previously~\cite{guo1, guo2} included pion
loop contributions by applying heavy meson chiral perturbation theory
at small $m_\pi$, following a series of works in this direction~\cite{lein}.
This leads to rapid,
nonanalytic variation when $m_\pi$ is smaller than about
400 - 500 MeV (which corresponds to a current quark mass $m_q \sim 60$ MeV).
When $m_\pi$ is larger than 400 - 500 MeV, the heavy meson mass
varies slowly and smoothly and is linear in $m_\pi^2$, as indicated by the
lattice data.   Such considerations of the behavior of the individual
meson masses led to the suggestion that the hyperfine splitting,
\mbox{$y_P=m_{P^*}-m_P$}, for the heavy meson, $P$,
can be extrapolated with the
following form:
\be
y_{P}=\bar{\sigma}_{P} + a_P +b_P m_\pi^2,
\label{a2}
\ee
where $\bar{\sigma}_P \equiv \sigma_{P^*}-\sigma_P$, and
$\sigma_{P(P^*)}$ the pion loop contribution to the meson mass of $P$
($P^*$).   This guarantees the correct chiral limit behavior, and $a_P$
and $b_P$ are fit parameters.   The resulting extrapolated
values for the hyperfine splittings for both $D$ and $B$ mesons are even
smaller than those obtained in the naive linear extrapolation~\cite{guo1,guo2}.

Here we investigate the appropriateness of the last two terms of \Eq{a2}
in representing the ``analytic background'' physics to which the chiral
loop contributions are added.   We investigate the constraints from two models:
the constituent quark potential model (CQM), and the Dyson-Schwinger
equation (DSE) model in ladder-rainbow truncation~\cite{Roberts:1994dr}.
For all the hadron
properties which have been calculated in lattice QCD, the lattice results vary
slowly and smoothly when $m_q$ is larger than about 60 MeV.   This is
characteristic of constituent quark behavior~\cite{cloet}  and suggests
the use of the constituent quark mass for an efficient description of hadron
properties in this region.   The CQM emphasizes such a concept.

On the other hand, the DSE ladder-rainbow model is a covariant modeling of
QCD that satisfies the chiral symmetry constraints.  In particular, it properly
incorporates dynamical chiral symmetry breaking~\cite{Maris:1997hd} which
plays a dominant role in hyperfine splitting of physical ground
states~\cite{Maris:2000zf,Tandy:2003hn}.
No explicit chiral loops are present at ladder-rainbow truncation; nevertheless
the relation between pseudoscalar mass and quark current mass has the correct
leading behavior at low and high quark mass~\cite{Ivanov:1998ms}.

An analytic background form for hyperfine splitting is deduced from the CQM in
Sec.~\ref{sec:potlmod},
and from the DSE model in Sec.~\ref{sec:DSEmodel} and comparisons are made there.
The chiral loop contributions to the hyperfine splitting are discussed in
Sec.~\ref{sec:piloops}, and the chirally extrapolated results are presented in
Sec.~\ref{sec:results}.   A summary is made in Sec.~\ref{sec:summary}.
\section{\label{sec:potlmod} Constituent Quark Potential Model}

The constituent quark model (CQM) has been shown to work quite well even
though it is a very simple model which has not been derived from Quantum
Chromodynamics. The constituent quark mass, $M_q$ (where $q$ denotes a
$u$ or $d$ quark), is
linked to the current quark mass, $m_q$, in the following way \cite{cloet}:
\be
M_q=M_\chi +c~m_q,
\label{a1}
\ee
where $c$ is of order 1 and $M_\chi$ is the constituent quark mass in the
chiral limit which originates from dynamical chiral symmetry breaking
in QCD.

In the constituent quark potential model, the potential between a quark
and an antiquark consists of scalar and vector parts.
The hyperfine splitting is caused by the relativistic spin-spin
interaction between the quark and the antiquark.
We consider the model where the vector part of the potential is
caused by one-gluon-exchange.   Then the hyperfine splitting for
the heavy meson $\bar Q q$ is~\cite{lucha}
\be
y_P^{\rm CQM} = \frac{32}{9}\pi \alpha_s \frac{1}{M_Q M_q} |\psi(0)|^2~~~,
\label{eq:cqm_split}
\ee
where $M_Q$ is the constituent quark mass of the heavy quark $Q$,
$\alpha_s$ is the strong coupling constant, and $\psi(0)$
is the wave function of the heavy meson at the origin.   On the other hand,
$\psi(0)$ can be expressed as~\cite{lucha}:
$|\psi(0)|^2=\frac{1}{2\pi} \frac{M_Q M_q}{M_Q + M_q}\langle V'(r)\rangle$.
In a heavy meson which consists of a light quark and a heavy antiquark the
dynamics is mainly determined by the light degrees of freedom due to
heavy quark symmetry.   The long distance part of $V(r)$ is the most relevant,
and for a typical potential model with linear confinement, $\langle V'(r)\rangle$
is essentially independent of quark masses.  Consequently, from the CQM we expect
the mass dependence of hyperfine splitting to be well-represented by
\be
y_P^{\rm CQM}=\frac{\tilde{d}_P}{M_q+M_Q}~~~,
\label{a4}
\ee
where $\tilde{d}_P$ is a constant.

The constituent quark mass $M_q$ in Eq. (\ref{a4}) will vary with
$m_\pi^2$ because of \Eq{a1} and the variation of current mass $m_q$ with $m_\pi^2$
due to chiral symmetry.   The latter variation we take from the
Gell-Mann--Oakes--Renner (GMOR) relation in the form:
$\frac{m_q}{m_q^{\rm phys}}=\frac{m_\pi^2}{(m_\pi^{\rm phys})^2}$,
where $m_q^{\rm phys}$ is the quark mass associated with the physical pion
mass, $m_\pi^{\rm phys}$.   Lattice studies indicate this relation is acceptable
up to $m_\pi \sim 1$ GeV .    Then with Eq. (\ref{a1}) one has
$M_q=M_\chi +\frac{c~m_q^{\rm phys}}{(m_\pi^{\rm phys})^2}m_\pi^2$
\cite{cloet}, and \Eq{a4} takes the more convenient form
\be
y_P^{\rm CQM}=\frac{d_P}{e_P+m_\pi^2}~~~.
\label{a7}
\ee
Here
$d_P \equiv \tilde{d}_P \frac{(m_\pi^{\rm phys})^2}{c~m_q^{\rm phys}}$
and $e_P \equiv e'+ M_Q \frac{(m_\pi^{\rm phys})^2}{c~m_q^{\rm phys}}$ with
$e'$ being defined as $M_\chi \frac{(m_\pi^{\rm phys})^2}{c~m_q^{\rm phys}}$.

We expect $y_P^{\rm CQM}$ to be a better representation of the
``analytic background'' physics than the last two terms of \Eq{a2}
for several reasons.   Firstly, while $y_P^{\rm CQM}$ shares with the background 
part of \Eq{a2} a linear behavior with 
small $m_\pi^2$, its dependence on $M_Q$ is consistent with heavy quark
effective theory~\cite{wise}. In contrast, \Eq{a2} provides no
explicit $M_Q$ dependence.   
Secondly, the two parameters of \Eq{a2} are totally
phenomenological while the parameter $e_P$ of $y_P^{\rm CQM}$ is
constrained by existing applications of the CQM to other baryon
physics. For example, the ratio of the constituent quark masses of
the strange quark and the $u,d$ quark is given by
\mbox{$\frac{M_s}{M_{u,d}}=$} \mbox{$\frac{e'+(m_\pi^2)_s}{e' +
(m_\pi^{\rm phys})^2}$}, where $(m_\pi^2)_s$ is the pion mass
squared when $m_q$ is equal to $m_s$. Use of the CQM and chiral
perturbation theory to guide the extrapolation of the magnetic
moments of the spin-1/2 baryon octet leads to \mbox{$M_s/M_{u,d}
\sim 1.3$}~\cite{cloet}.   We let this ratio vary between 1.2 and
1.4.    Use of the GMOR relation between $m_\pi^2$ and $m_q$ leads
to \mbox{$(m_\pi^2)_s \sim 0.5$ GeV$^2$}.   Consequently we allow
the range of $e'$ to be from 1.2 GeV$^2$ to 2.5 GeV$^2$. The CQM
fit in \Ref{cloet} found \mbox{$c~m_q^{\rm phys} \sim 5.90$} MeV.
If we take the heavy constituent masses to be \mbox{$M_c \sim
1.3-1.7$}~GeV and \mbox{$M_b \sim 4.6-5.0$}~GeV, then we have the
parameter ranges \mbox{$e_D \sim$} \mbox{$5.5-8.2~{\rm GeV}^2$}
and \mbox{$e_B \sim$} \mbox{$16.5-19.2~{\rm GeV}^2$}.

In Sec.~\ref{sec:results} we add the pion loop contribution to $y_P^{\rm CQM}$
and fit the single free parameter $d_P$ to the lattice data, while constraining
$e_P$ to be within the above expected range.   For the sake
of comparison, the $m_\pi^2$ dependence from $y_P^{\rm CQM}$ alone is
illustrated here by a preliminary fit of $y_P^{\rm CQM}$ to the lattice data,
allowing both $d_P$ and $e_P$ to vary freely.
This is displayed in \Figs{fig:Dsplit_DSE} and \ref{fig:Bsplit_DSE}.
(The meaning of the model denoted CQM$^\prime$ in those figures is explained
in the next Section.)  We obtain
\mbox{$d_D =$} \mbox{$0.547~{\rm GeV}^3$}, \mbox{$e_D=$} \mbox{$4.462~{\rm GeV}^2$},
and \mbox{$d_B =$} \mbox{$0.447~{\rm GeV}^3$}, \mbox{$e_B=$}
\mbox{$14.71~{\rm GeV}^2$} for $y_D^{\rm CQM}$ and  $y_B^{\rm CQM}$
respectively.  These $e_P$ values are just below the expected
range.
\section{\label{sec:DSEmodel} Dyson-Schwinger Equation Model}

Information on hyperfine splitting in the absence of chiral loops
can also be obtained from the modeling of mesons through the Dyson-Schwinger
equations (DSE) of QCD in rainbow-ladder truncation~\cite{Maris:1999nt,Tandy:2003hn}.
This a Poincar\'e covariant approach that implements the correct one-loop
renormalization group behavior of QCD~\cite{Maris:1997tm} and  the
relevant Ward--Takahashi identities.   The resulting axial current
conservation preserves the Goldstone nature of the pseudoscalars~\cite{Maris:1997hd}.
With one infrared parameter (besides the
quark current masses) the model~\cite{Maris:1999nt} provides an efficient
description of the masses and decay constants of the light-quark
pseudoscalar and vector mesons~\cite{Maris:1997tm,Maris:1999nt},  the elastic
charge form factors $F_\pi(Q^2)$ and $F_K(Q^2)$~\cite{Maris:2000sk} and
the electroweak transition form
factors of the pseudoscalars and vectors~\cite{Maris:2002mz,Ji:2001pj}.
\begin{figure}[hbt]
\centering{\epsfig{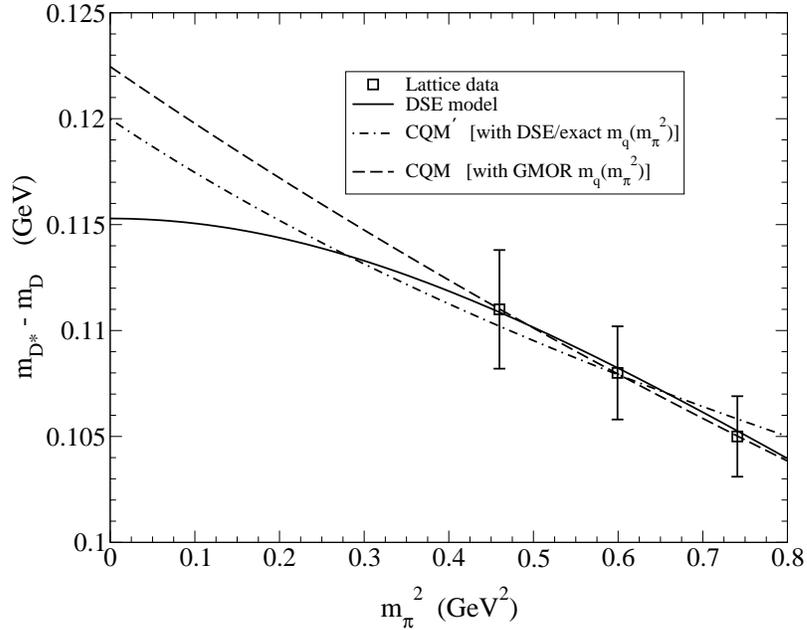}}
\caption[]{The ``analytic background'' contribution to the $D^*-D$ mass splitting
as obtained from the ladder-rainbow DSE model and from the CQM model.   They are fitted
to the lattice data for comparison purposes.  }
\label{fig:Dsplit_DSE}
\end{figure}

The current quark mass dependence of the DSE results satisfies the exact QCD
pseudoscalar mass relation~\cite{Maris:1997hd}.
For $qQ$ mesons it takes the form~\cite{Maris:1997hd}
\mbox{$m^2_P \; f_P =$} \mbox{$(m_q + m_Q)\; R_P$}
where the electroweak decay constant $f_P$, and the projection $R_P$ of the
Bethe-Salpeter wave function onto $\gamma_5$ at the origin of $\bar q q$
separation, are dependent upon the current masses $m_q$ and $m_Q$.
The quark masses and $R_P$ are dependent upon renormalization scale $\mu$
but the mass relation is not.  The chiral limit of $R_P$ is~\cite{Maris:1997hd}
\mbox{$R_{P,\mu}^0 = -\langle \bar q q \rangle_\mu/f_P^0$}, where $f_P^0$
is the chiral limit value; thus the GMOR relation follows as a collorary
of the exact relation in the low mass limit.  [The error in the GMOR relation
at the $K$ meson is 4\%, at $m_Q= 0.4$~GeV the error is 14\%, while at the
D meson the error is 30\%~\cite{Tandy:2003hn}.]   The heavy quark limit of the
relation produces~\cite{Ivanov:1998ms} \mbox{$m_P \propto m_Q$} in accord with
constituent quark behavior.

The current quark mass dependence of the DSE results for the ground state
$m_P$ and $m_{P^*}$ has been studied recently~\cite{Maris:2000zf,Tandy:2003hn}
with $m_q$ fixed at $m_u$ while $m_Q$ varies in the range 0--300 MeV.
The results are summarized well in terms of a single parameter $\alpha$ by the forms
\ba
\label{eq:ps_DSEfit}
m_P^2 &=& \alpha_0~(m_q + m_Q) + \alpha^2~(m_q + m_Q)^2 \\
m_{P^*} &=& m_\rho + \alpha~(m_q + m_Q - 2 m_u)~~~,
\label{eq:v_DSEfit}
\ea
where current quark masses are defined at a scale of \mbox{$\mu = 1$}~GeV.
The large \mbox{$m_q=m_Q$} behavior is consistent with recent DSE model
studies~\cite{Ivanov:1998ms,Bhagwat:2004hn}.
Here \mbox{$\alpha_0 = 1.724$}~GeV is the DSE model-exact value of
$-\langle \bar q q \rangle^0_\mu/(f_P^0)^2$; use of this
term alone corresponds to the GMOR relation.  The $\alpha^2$ term accounts
for the correction required by the exact pseudoscalar mass relation as
reflected in the DSE results.  The value \mbox{$\alpha=1.079$} provides a
good overall representation of the $\pi/\rho, K/K^*, D/D^*$ and $B/B^*$ meson
masses.   The absence of a linear term in \Eq{eq:ps_DSEfit} respects dynamical 
chiral symmetry breaking and this in turn provides a good representation of
the increase in the physical hyperfine splittings with decreasing mass.

\begin{figure}[hbt]
\centering{\epsfig{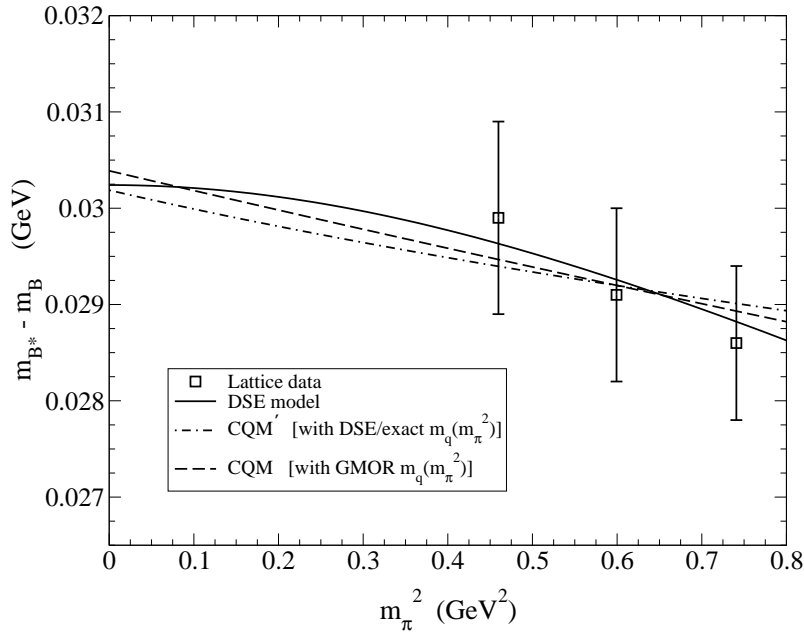}}
\caption[]{The ``analytic background'' contribution to the $B^*-B$ mass splitting
as obtained from the ladder-rainbow DSE model and from the CQM model.   They
are fitted to the lattice data for comparison purposes.  }
\label{fig:Bsplit_DSE}
\end{figure}

For any value of the parameter $\alpha$, the hyperfine splitting obtained
from \Eqs{eq:ps_DSEfit} and (\ref{eq:v_DSEfit}) in the form
\mbox{$y_P^{\rm DSE} = m_{P^*} - m_P$} becomes independent of $m_Q$ as
\mbox{$m_Q \to \infty$}.  In that limit, heavy quark effective theory
makes a more specific statement, namely \mbox{$y_P \to 0$}.   This can
be satisfied with the special value
\be
\alpha = \frac{m_\rho}{4 m_u}\; \big\{ 1 -
\sqrt{1 - \frac{4 m_u \alpha_0}{m_\rho^2} } \big\}~~~,
\label{eq:specalpha}
\ee
which yields \mbox{$\alpha =1.138$}.
Since the employed DSE calculations are limited by \mbox{$m_Q \leq
300~{\rm MeV}$}, we prefer to use \mbox{$\alpha =1.138$} as a way
of incorporating large $m_Q$ physics.  The ``analytic background''
as deduced from the DSE model is thus fully proscribed in
magnitude and $m_\pi$ dependence.   To make an initial assessment
of this $m_\pi$-dependent background, in comparison to $y^{\rm
CQM}_P$, we fit the form \be y_P^{\rm DSE} = m_{P^*} - m_P +
\Delta_P~~~, \label{eq:yDSE} \ee to the lattice data with
$\Delta_P$ being a free constant.  The results are shown in
\Figs{fig:Dsplit_DSE} and \ref{fig:Bsplit_DSE}.   The required
shifts \mbox{$\Delta_D = -0.0244$}~GeV and \mbox{$\Delta_B =
-0.0218$}~GeV perform a role which is later taken over by the
$m_\pi$-dependent pion loop contributions.   As we note later, the
pion loop contribution in the vicinity of the lattice data is
indeed negative and of this order.

We note that the conversion of $m_q$-dependence into $m_\pi$-dependence,
consistently within the DSE, has required the inversion of  \Eq{eq:ps_DSEfit}
with \mbox{$m_Q=m_q$}.    One may ask whether the resulting
$m_q^{\rm DSE}(m_\pi^2)$ differs sufficiently from the GMOR result
$m_q^{\rm GMOR}(m_\pi^2)$ to warrant concern for the sensitive
hyperfine splittings considered here.  We provide information on this
in \Figs{fig:Dsplit_DSE} and \ref{fig:Bsplit_DSE} by means of two versions
of the $y_P^{\rm CQM}$ description of the background.   The version labelled CQM
employs $m_q^{\rm GMOR}(m_\pi^2)$; the version labelled CQM$^\prime$ employs
$m_q^{\rm DSE}(m_\pi^2)$.    The results show this not to be a significant source
of uncertaintity; it amounts to a maximum of only 3~MeV in $y_D$ at
\mbox{$m_\pi = 0$}, and less for $y_B$.    Uncertainties in the DSE
description of the background from the parameterization in
\Eqs{eq:ps_DSEfit} and (\ref{eq:v_DSEfit}) are not large.  If we were to retain
the parameter value \mbox{$\alpha=1.079$} which is preferred by
the low mass DSE calculations, the changes to the results shown in
\Figs{fig:Dsplit_DSE} and \ref{fig:Bsplit_DSE} are insignificant.

From \Figs{fig:Dsplit_DSE} and \ref{fig:Bsplit_DSE} it is evident that there
is more curvature in $y_P^{\rm DSE}$ than in $y_P^{\rm CQM}$.
This is due to the non-linearity imposed by chiral symmetry; the DSE model
respects the axial vector Ward-Takahashi identity~\cite{Maris:1997hd}.
In Table~\ref{table0} we summarize these representations of the ``analytic
background'' in terms of would-be contributions to the hyperfine
splittings at \mbox{$m_\pi = 0$}; we also include the linear extrapolation results.
The DSE model and the CQM are consistent with each other to about 5\% and both
tend to be slightly below the naive linear extrapolation.  However, the linearly
extrapolated splittings are already significantly too low: 15\% low in the D
meson case and 30\% low for the B meson.

\begin{table}[hbt]
\beforetab
\begin{tabular}{cccccc}
\firsthline

& CQM &  CQM$^\prime$ & DSE & Linear & Expt \\
\midhline
$y_D$  & 0.1224  & 0.120   & 0.1153  & 0.120  & 0.140   \\

$y_B$  & 0.0304  & 0.0302   & 0.0303  & 0.032  & 0.046   \\
\lasthline
\end{tabular}
\aftertab
\captionaftertab[]{\label{table0} Various background models extrapolated
to \mbox{$m_\pi=0$} and expressed as contributions to the hyperfine splitting.
As explained in the text, $m_q^{\rm GMOR}(m_\pi^2)$ is used in  CQM, while
$m_q^{\rm DSE}(m_\pi^2)$ is used in CQM$^\prime$.   All quantities are in GeV
units. }
\end{table}
\section{\label{sec:piloops} Pion Loops}

\begin{figure}[ht]
\centering{\epsfig{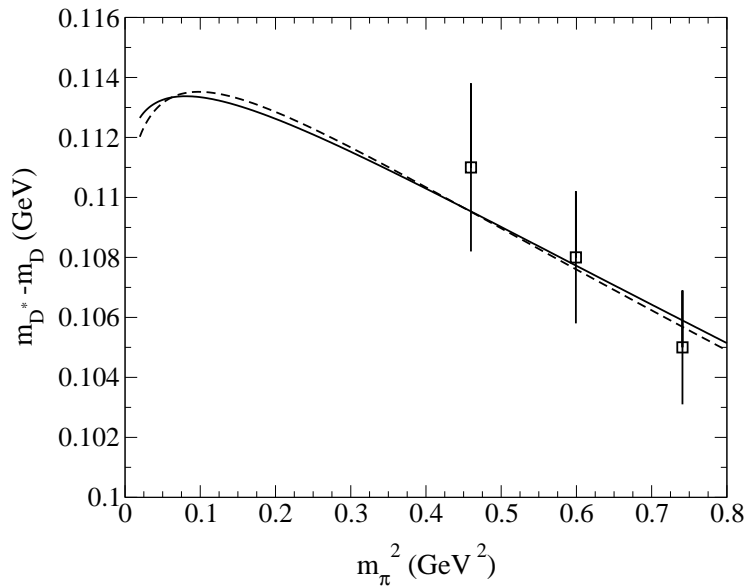}}
\caption{Extrapolation of the lattice data for the $D^*-D$ hyperfine
splitting using chiral loops, with a dipole form factor (solid line) and with
a sharp-cutoff form factor (dashed line), added to the CQM model.
The parameters are: $\Lambda$=0.4 GeV, $\lambda_2=-0.015$ GeV$^2$,
$g^2=0.4$ and $e_D$=6.9 GeV$^2$.}
\label{fig:lac}
\end{figure}

The pion loop contributions to the masses of heavy vector and heavy
pseudoscalar mesons were calculated in Ref. \cite{guo1}. The interaction
between the pion and heavy mesons is generated from the following
chiral Lagrangian
for heavy mesons which is invariant under both chiral symmetry
and heavy quark symmetry in the chiral and heavy quark limits, respectively
\cite{hlcpt1}:
\begin{equation}
{\cal L}= -{\rm Tr}[\bar{H}_a iv_\mu (D^\mu H)_a]+g {\rm Tr}(\bar{H}_a
H_b \gamma_\mu A^\mu_{ba} \gamma_5),
\label{a9}
\end{equation}
where $D^\mu$ is the covariant derivative and $A^\mu$ is the axial-vector
field, both of which include Goldstone boson fields,
$g$ is the coupling constant describing the interactions between
heavy mesons and Goldstone bosons, and $H_a$ ($a$=1, 2, 3 for $u$, $d$, $s$
quarks, respectively) is a field operator which includes heavy pseudoscalar
($P$) and heavy vector ($P^*$) mesons: $H_a=\frac{1+\rlap/v}{2}
(P_a^{* \mu}\gamma_\mu-P_a \gamma_5)$ ($v$ is the velocity of the heavy meson).
The correction term $\frac{\lambda_2}{m_Q} {\rm Tr} \bar{H}_a \sigma^{\mu\nu}
H_a \sigma_{\mu\nu}$ was added to Eq. (\ref{a9}) (where $\lambda_2$ is
a parameter) since this term is responsible for hyperfine splittings.
Based on experimental data we chose $g^2$ to vary between 0.3 and 0.5
and $\lambda_2$ between -0.03 GeV$^2$ and -0.02~GeV$^2$ \cite{guo1}.
With Eq. (\ref{a9}) $\bar{\sigma}_P$ can be expressed as the following:
\begin{eqnarray}
\bar{\sigma}_P &=&-\frac{g^2}{8\pi^2 f_\pi^2}\int_0^\infty {\rm d}k
\frac{k^4u^2(k)}{\sqrt{k^2+m_\pi^2}(\sqrt{k^2+m_\pi^2}-\Delta)}\nn \\
&-&\frac{g^2}{4\pi^2 f_\pi^2}\int_0^\infty {\rm d}k
\frac{k^4u^2(k)}{k^2+m_\pi^2}\nn \\
&+&\frac{3g^2}{8\pi^2 f_\pi^2}\int_0^\infty {\rm d}k
\frac{k^4u^2(k)}{\sqrt{k^2+m_\pi^2}(\sqrt{k^2+m_\pi^2}+\Delta)},
\label{a10}
\end{eqnarray}
where $f_\pi=132$ MeV is the pion decay constant,
$\Delta= -8\lambda_2/m_Q$, $k$ is the absolute value of the three-momentum
of the pion in the loop,
and $u(k)$ is an ultra-violet regulator. Since the leading
nonanalytic contribution of pion loops is only associated with the infrared
behavior of the integrals in Eq. (\ref{a10}), it does not
depend on the details of the regulator. In this work we choose two different
forms for the regulator. One is the sharp-cutoff, $\theta(\Lambda - k)$,
and the other the dipole form, which is more realistic,
$\Lambda^4/(\Lambda^2 + k^2)^2$. Here $\Lambda$
characterizes the finite size of the source of the pion - i.e, the heavy
meson's radius. In the fit we let $\Lambda$ vary between 0.4 GeV and
0.6 GeV as in our previous work.

The hyperfine splittings for $D$ and $B$ mesons were calculated in
Ref. \cite{hein}, where NRQCD was used to treat the heavy quarks.
For the bare gauge coupling $\beta=5.7$, the inverse lattice
spacing $a^{-1}$ is about 1.116 GeV. The box size is 2.1 fm, corresponding to
the volume $12^3 \times 24$. In their simulations, three different values
for the hopping parameter $\kappa$, 0.1380, 0.1390, and 0.1400, were used.
The light quark mass is related to $\kappa$ as
$m_q=\frac{1}{2a}(1/\kappa -1/\kappa_c)$, with $\kappa_c=0.1434$.
\begin{figure}[ht]
\centering{\epsfig{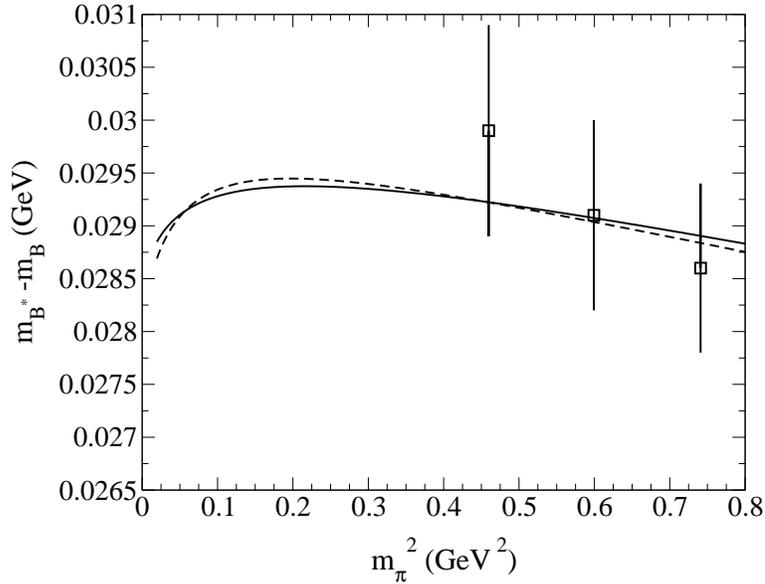}}
\caption{Extrapolation of the lattice data for the $B^*-B$ hyperfine
splitting using chiral loops, with a dipole form factor (solid line) and with
a sharp-cutoff form factor (dashed line), added to the CQM model.
The parameters are the same as those in Fig. \ref{fig:lac}
except that $e_B$=17.9 GeV$^2$.}
\label{fig:lab}
\end{figure}

The next step is to fit these lattice data with the form of \Eq{eq:extrap}
to determine the parameter $d$.
The formulas in Eq. (\ref{a10}) are obtained in the infinite volume limit.
Since the lattice simulations are performed on a finite volume grid, the
finite size effects should be taken into account.
Following the second reference in Ref. \cite{lein} and Ref. \cite{young} we
replace the continuum self-energy integral in Eq. (\ref{a10}) by a sum over
the discrete pion momenta which are allowed on the lattice:
$\int {\rm d}^3 k \approx \left(\frac{2\pi}{aL}\right)^3
\sum_{ k_x, k_y, k_z}$,
where the discrete momenta $k_x, k_y, k_z$ are given by $\frac{2\pi n}{aL}$,
$L$ is the number of lattice sites in each spatial direction,
and the integer $n$ satifies the constraint $-\frac{L}{2} < n \le \frac{L}{2}$.
With $1/a=1.116$ GeV and $L=12$, the smallest momentum allowed on the lattice,
$2\pi /a L$, equals 0.58 GeV.

With the least square fitting method we fix the parameter $d_P$ for
different values of $e_P$, $\Lambda$, $g$, and $\lambda_2$
in both the sharp-cutoff and
the dipole schemes. It is noted that when $m_\pi^2 \leq \Delta^2$
there is a pole in the first integral in Eq. (\ref{a10}). In this case, we
have kept the principal value of the integral which is real. The difference
between the integral and its principal value is
imaginary and corresponds to the width of the heavy meson.
\section{\label{sec:results} Results for Chiral Extrapolation}

Based on the above considerations, we extrapolate the lattice data with
the form
\be
y_P = \bar{\sigma}_P + \frac{d_P}{e_P+m_\pi^2}~~~.
\label{eq:extrap}
\ee
$d_P$ is a free parameter while, from Sec.~\ref{sec:potlmod}, we constrain
$e_P$ to be within \mbox{$e_D \sim$} \mbox{$5.5-8.2~{\rm GeV}^2$} and
\mbox{$e_B \sim$} \mbox{$16.5-19.2~{\rm GeV}^2$}.
This direct extrapolation of the hyperfine splitting should lead to
smaller errors than the separate extrapolation of masses carried out in \Ref{guo1}.

For $\bar{\sigma}_P$, the allowed ranges of the parameters $\Lambda$, $g$, and
$\lambda_2$ are those described in Sec~\ref{sec:piloops}.  In \Figs{fig:lac} and
\ref{fig:lab} we show the fits to the lattice data  for $\Lambda=0.4$ GeV and
for the intermediate values of
other parameters in both the dipole and the sharp-cutoff
regulator schemes.    When \mbox{$m_\pi <$} 500~MeV the extrapolations start
to deviate significantly from the constituent quark model.  We require the fits
to be within the uncertainties of the lattice data; this is always possible for
$\Lambda=0.4$ GeV but not for $\Lambda=0.5$ GeV or
0.6 GeV$^{\rm\footnotemark[1]}$\footnotetext[1]{When the fit results are
not required to be within the lattice uncertainties the high limiting
values of $y_P$ remain unchanged while
the low limiting values are reduced by about 8$\sim$17\%. The values of
the parameters $d_P$, $a_P$, and $b_P$ are not affected.}.
In Table~\ref{table1} the hyperfine splittings $y_P$, extrapolated to the physical
pion mass, are listed as a range of values allowed by the variation of parameters 
as described above.  We also give separately the relative uncertainties $\Delta y/y$ 
caused by the errors in the
lattice data.   To assess the influence from the description of the background,
we show results from use of the CQM background, \Eq{eq:extrap}, and from the
linear background used in previous work, \Eq{a2}.    The associated parameters
are given in the Table.

\begin{table}[htb]
\beforetab
\begin{tabular}{ccccc}
\firsthline
    & \multicolumn{2}{c}{$D$ Mesons} & \multicolumn{2}{c}{$B$ Mesons}\\
\midhline
Expt (GeV) & \multicolumn{2}{c}{0.140} & \multicolumn{2}{c}{0.046} \\
\midhline
Form Factor            & Dipole   & Sharp  & Dipole   &  Sharp   \\
\midhline
With CQM background    &          &         &          &           \\
  &  &  &  &  \\
$y_P$ (GeV)      & 0.0983-0.1150  &  0.0978-0.1143   &  0.0260-0.0289 &  0.0259-0.0288  \\
$d_P$ (GeV$^3$)  & 0.6675-1.0601  &  0.6590-1.0237   &  0.5048-0.6467 &  0.4982-0.6245  \\
$\Delta y /y$ (\%) & 1.24-1.45    &  1.25-1.46       &  1.82-2.02     &  1.83-2.03      \\
  &  &  &  &  \\
\midhline
With Linear background &          &                  &                &                 \\
  &  &  &  &  \\
$y_P$ (GeV)     &   0.1000-0.1162 & 0.0965-0.1149    &  0.0270-0.0309 &  0.0256-0.0306   \\
$a_P$ (GeV)     &  0.1226-0.1395  & 0.1203-0.1375    &  0.0326-0.0374 &  0.0320-0.0369    \\
$-b_P$ (GeV$^{-1}$) & 0.0308-0.0216 & 0.0340-0.0203  & 0.0077-0.0051   &  0.0086-0.0047    \\
$\Delta y /y$ (\%) & 6.28-7.30     & 6.36-7.57       & 8.90-10.18      & 9.00-10.76       \\
  &  &  &  &  \\
\lasthline
\end{tabular}
\aftertab
\captionaftertab[]{\label{table1} The hyperfine splittings extrapolated to the physical
pion mass through chiral loops added to the background described via the CQM
(with parameter $d_P$) or the linear representation (with parameters  $a_P$ and $b_P$).
Results arising from the dipole form factor and the sharp-cutoff form factor
are shown.    The splittings $y_P$ are shown as a range of values allowed by the variation
of parameters as discussed in the text.   Separately given are the relative uncertainties 
$\Delta y/y$ caused by the errors in the lattice data.
}
\end{table}

With our parameter ranges,  the extrapolated hyperfine splitting with CQM
background
for $D$ mesons varies from 0.0983 GeV to 0.115 GeV with the dipole form
factor, and  from 0.0978 GeV to 0.114 GeV with the sharp-cutoff form factor.
For $B$ mesons, it varies from 0.0260 GeV to 0.0289 GeV and from
0.0259 GeV to 0.0288 GeV, respectively.   On the other hand,
with use of the linear background the $D$ meson result is in the range
0.100-0.116 GeV from the dipole form factor, and in the range 0.0965-0.115 GeV
from the sharp-cutoff form factor.
For $B$ mesons, the extrapolated hyperfine splittings are 0.0270-0.0309 GeV and
0.0256-0.0306 GeV, respectively.   The largest extrapolated hyperfine
splittings we could obtain are about 115 MeV and 29 MeV for $D$ and $B$ mesons,
respectively from the CQM background; with the linear background,
these become 116 MeV and 31 MeV, respectively.   The present approach is unable
to do better than 17\% below experiment for the $D$ splitting and 35\% below
experiment for the $B$ splitting.   These are further from experiment than the
naive linear extrapolation which, as described in Sec.~\ref{sec:DSEmodel}, 
produces a D splitting 15\% below and a B splitting 30\% below.

One can compare the values at the midpoints of the ranges of the CQM background
parameters $e_P$ and $d_P$  that fit the lattice data in the presence
of the pion loop contribution, to the corresponding values obtained
in Sec.~\ref{sec:potlmod} by the fit of the CQM background alone to the data.
This confirms that at \mbox{$m_\pi^2 \approx 0.6~{\rm GeV}^2$} the pion loop
contribution to $y_D$ is typically $-10$~MeV, and the contribution to $y_B$ is
typically $-20$~MeV.  These values are of the same order as the typical shifts
$\Delta_D$ and $\Delta_B$ found in Sec.~\ref{sec:DSEmodel} to align the DSE
background with the lattice data.

In addition to the uncertainties which are associated with the parameters
$\lambda_2$, $g$, $\Lambda$, and $e_P$ we also analysed the uncertainties
which are caused by the errors in the lattice data.   As expected, these
uncertainties are much smaller (less than 2\%) with the improved CQM background
description compared to the linear treatment of the background (6$\sim$11\%).
Taking into account such uncertainties, we can see that
the ranges of the extrapolated hyperfine splittings produced with the different
background treatments are compatible with each other.

Another interesting quantity is the slope of the fit line at large $m_\pi$.
From \Figs{fig:lac} and \ref{fig:lab} we can see that the slope of the fit for
$B$ mesons is much smaller than that for $D$ mesons.   The CQM representation
of the background carries a $1/M_Q^2$ dependence for the slope and provides
a natural fit.  In contrast, the linear representation of the background
contains no $M_Q$ dependence and this contributes to the larger uncertainties
that result.    The difference of slopes in these two extrapolation forms could be
tested with more accurate lattice data.

In the above fit we have treated the lattice data as truly statistical.
Since the lattice data are highly correlated we could also fit the top
and the bottom of the three data points in \Figs{fig:lac} and
\ref{fig:lab} to give the
ranges of the fit. This may lead to a little change in the results.
As an example, we consider the largest extrapolated hyperfine splittings for
$D$ and $B$ mesons in Figures \ref{fig:lac} and \ref{fig:lab}. If we fit the
top of the three lattice data in these two figures in the dipole scheme
we obtain the extrapolated results as 0.1177 GeV and 0.0299 GeV for $D$ and $B$
mesons, respectively. These values are a little bigger than those obtained
when we treat the lattice data as truly statistical.

\section{\label{sec:summary} Summary}

We have explored a number of issues that arise in the chiral
extrapolation of the lattice data for hyperfine splittings of $D$ and $B$ mesons
from the domain of large mass of the light quark to the physical point.
We improve on previous work by investigating the description of the non-chiral
background via both the constituent quark model (CQM) and the modeling of QCD
via the Dyson Schwinger equations (DSEs) in ladder-rainbow truncation.  We
find that these descriptions of the background are consistent with each other.
Thus the form employed here for the analytic background for extrapolation in
$m_\pi^2$ is consistent with: chiral symmetry, heavy quark symmetry in the 
limit of infinite heavy quark mass, the constituent quark model, and
the DSE modeling of QCD.
We adopt the CQM background, add the difference of the chiral loop self-energies
for vector and pseudoscalar mesons, and fit to the lattice data. We extract a
range of extrapolated hyperfine splittings at the physical pion mass for
both $D$ and $B$ mesons, as allowed by the theoretical uncertainties in 
parameters and pion-baryon form factors.   

In comparison to our previous work~\cite{guo1, guo2} that employed an analytic 
background linear in $m_\pi^2$, the present approach has a better physical 
basis in the dependence of the background upon both $M_Q$ and $m_\pi^2$.
The pion-loop contribution $\bar{\sigma}_P$ produces essentially all of the curvature 
evident in \Figs{fig:lac} and \ref{fig:lab} at low $m_\pi^2$ and this
is clearly less curvature than what we obtained in Ref.~\cite{guo1}.  
This is because the present approach deals directly with the hyperfine splitting
and its extrapolation.   There is a single form factor
range $\Lambda$ for a vector-pseudoscalar meson pair.   In contrast, the prevous 
work~\cite{guo1} dealt with extrapolation of the meson masses separately
before the difference is formed.   That procedure employed separate form factor 
parameters $\Lambda$ for each 
meson and the resulting splittings were a difference of two quantities having
different curvatures at low $m_\pi^2$.    The present more direct procedure
has less curvature as a result of being better constrained by physics.   It has 
smaller uncertainties in the extrapolated splittings from errors in the lattice 
data.   We conclude that the present procedure is more reliable.  

In contrast to extrapolations of lattice data for hadron 
masses~\cite{Leinweber:2003dg} and magnetic moments~\cite{magm_extrap}, 
the 1-loop chiral loop self-energies do not
improve the agreement of the extrapolated results for hyperfine
splittings in comparison with experiment.   This work has explored the 
analytic background carefully and we conclude that this is not a likely 
explanation.    Although 
sub-leading nonanalytic behavior from higher order chiral loops
could be examined, we conclude that more reliable 
lattice simulations for hyperfine splittings are required to resolve this
situation.    One would like to have data 
for full QCD, so as to not rely on the quenched approximation.
The experimental mass splittings are small: 7\% for
D mesons and 0.8\% for B mesons.  The ${\bf
\sigma\cdot B}$ term in NRQCD is of order $1/M_Q$ and small corrections can
have a magnified impact.  The coefficient of this term may well be increased by the
inclusion of radiative corrections beyond tadpole improvement, the possible 
light quark mass dependence of the clover coefficient in the clover action for 
light quarks, and higher order terms in NRQCD.
With the physical
improvements incorporated here, the extrapolation is unable to do
better than 17\% below experiment for the $D$ splitting and 35\%
below experiment for the $B$ splitting.   These are further from
experiment than the naive linear extrapolation which produces a D
splitting 15\% below and a B splitting 30\% below.   The fact that the deviation
from experiment for $B$ splitting is consistently double that for $D$ 
splitting is suggestive of  systematic errors in the lattice data. 
Higher quality lattice data would stimulate greater
theoretical scrutiny of both analytic and nonanalytic behavior in
$m_\pi^2$ for such a sensitive quantity. 

\begin{acknowledge}
Craig Roberts is acknowledged for helpful conversations.
This work was supported by the Australian Research Council, by NSF
grants No. PHY-0071361, INT-0129236, and PHY-0301190, by DOE
contract DE-AC05-84ER40150, under which SURA operates Jefferson
Lab, and by the Special Grants for "Jing Shi Scholar" of Beijing
Normal University.
\end{acknowledge}


\end{document}